\documentclass[pra,twocolumn,showpacs,superscriptaddress,floatfix]{revtex4-2}
\usepackage{graphicx}
\usepackage{epstopdf} 
\usepackage{amsmath}
\usepackage{epsfig}
\usepackage{latexsym}
\usepackage{amsfonts}
\usepackage{graphics}
\usepackage{bm}
\usepackage{braket}
\usepackage{dsfont}
\usepackage{soul}
\usepackage{ulem}
\usepackage{bbm}
\usepackage{hyperref}

\usepackage{mathtools}

\usepackage{helvet}

\graphicspath{ {./Graphics/} }
\begin{document}

\date{\today}
\author{J. Mumford}
\affiliation{Department of Physics and Astronomy, McMaster University, 1280 Main St.\ W., Hamilton, ON, L8S 4M1, Canada}
\affiliation{Homer L. Dodge Department of Physics and Astronomy, The University of Oklahoma, Norman, OK 73019, USA}
\affiliation{Center for Quantum Research and Technology, The University of Oklahoma, Norman, OK 73019, USA}
\author{D. Kamp}
\affiliation{Department of Physics and Astronomy, McMaster University, 1280 Main St.\ W., Hamilton, ON, L8S 4M1, Canada}
\author{D. H. J. O'Dell}
\affiliation{Department of Physics and Astronomy, McMaster University, 1280 Main St.\ W., Hamilton, ON, L8S 4M1, Canada}

\title{Gauge potentials and vortices in the Fock space of a pair of periodically driven Bose-Einstein condensates}

\begin{abstract}
We perform a theoretical study of the coupled dynamics of two species of Bose-Einstein condensates (BECs) in a double-well potential where both the tunneling and the interatomic interactions are driven periodically in time. The population difference between the wells of each species  gives rise to a two-dimensional lattice in Fock space with dimensions given by the number of atoms in each BEC.   We use a Floquet analysis to derive an effective Hamiltonian that acts in this Fock space and find that it contains an artificial gauge field. This system simulates noninteracting particles in a tight-binding lattice subject to an additional harmonic potential and vector potential. When the intra-species interactions are attractive there is a critical value at which the ground state undergoes a transition from a Gaussian state to a quantized vortex state in Fock space.   The transition can be quantified in terms of the angular momentum and the entanglement entropy of the ground state with both showing sudden jumps as the intra-species interactions become stronger.  The stability of the vortex state vanishes in the thermodynamic limit.
\end{abstract}


\maketitle

\section{\label{Sec:Intro}Introduction}
Synthetic dimensions are artificial extra dimensions that mimic  real spatial ones.  They considerably expand the range of complex quantum systems that can be simulated using simpler, more easily controlled platforms such as ultracold atoms \cite{boada12,ozawa19b}, photons \cite{ozawa19a,smirnova20} and trapped ions \cite{blatt12,barredo18}, with examples of the internal and external degrees of freedom that can be used to make synthetic dimensions including harmonic oscillator \cite{price17}, spin \cite{celi14,mancini15,stuhl15,anisimovas16}, momentum \cite{an17,meier16,xie19}, rotational \cite{flob15}, Rydberg \cite{leseleuc19,kanungo22} and photon resonator \cite{hafezi13,mittal16} states.

Another useful tool for quantum simulation is provided by artificial gauge potentials. These mimic the action of electric and magnetic fields on charges irrespective of whether the underlying physical particles are charged or not \cite{lin09,lin11,jimenez12,dalibard11}. For ultracold atoms, a popular experimental configuration is to load them into an optical lattice which can then be periodically modulated to generate effective gauge potentials \cite{aidelsburger11,struck11,aidelsburger13,miyake13,jotzu14,goldman14a,goldman14b,eckardt15,bukov15}. The ability to simulate gauge fields in a controlled and tuneable manner has led to new proposals for investigating  topological phases \cite{goldman16, creffield16,zhang18,aidelsburger18,cooper} and exotic phases of matter \cite{khemani16,else16}, as well as experiments studying quantum Hall physics on entirely different physical platforms to their original settings \cite{stuhl15,chalopin20}. Artificial gauge fields not only provide a versatile experimental method for probing fundamental aspects of quantum mechanics and condensed matter physics but also enable new technological applications in quantum information processing and quantum simulation \cite{jaksch05,bloch08,altman21}.

Fock states provide yet another possible set of states that can be used as building blocks for synthetic dimensions. Unlike the more traditionally used internal and external states listed above, Fock states are defined via the occupation numbers of particles in the different available modes  leading to the concept of a Fock-state lattice (FSL) \cite{weiwang16,saugmann22}. Systems with a large number of particles and a limited number of modes are particularly well-suited for constructing large synthetic dimensions in this way because each mode acts as a single synthetic dimension, and the number of particles that can occupy a mode serves as the length of the dimension.   In this context, particles with bosonic statistics are preferable due to their ability to occupy the same mode simultaneously.  One example of a FSL is provided by photon states in optical cavities; when two or three optical cavities are coupled via a two-level atom such a FSL can realize the Su-Schrieffer-Heeger (SSH) model or a Lifshitz  topological phase transition between a semimetal and a three band insulator, respectively \cite{weiwang16,cai21}. Likewise, experiments with superconducting circuits have successfully constructed one- and two-dimensional (2D) FSLs  which can mimic both the SSH and Haldane models \cite{deng22}.  Of particular relevance to the current paper is a recent proposal by one of us (JM) based on two species of atomic Bose-Einstein condensates (BECs) confined in a double-well potential which realizes a two-dimensional FSL \cite{mumford22a}. This system can simulate strong synthetic gauge fields by intense periodic driving of the interactions between the two species of atoms leading to topological behavior in the form of chiral edge states.

Ever since the pioneering work by Onsager \cite{onsager49} and Feynman \cite{feynman55},  quantized vortices have been regarded as the quintessential example of topological excitations. The original system of interest was liquid helium II, and subsequent experiments by Hall and Vinen (see \cite{vinen1956,vinen1961}) confirmed that the circulation $\oint \mathbf{v}_{s} \cdot d \mathbf{r} $ is quantized in units of $h/m$, where $h$ is Planck's constant, $m$ is the mass of a helium atom and $\mathbf{v}_{s}$ is the velocity of the superfluid component. Shortly afterwards it was  realized that quantized vortices also form in type II superconductors in the presence of magnetic fields \cite{abrikosov1957,essmann1967}. In this latter case the circulating current is charged and each vortex encloses a quantum of flux $h/2e$ where $e$ is the electron charge. More recently, quantized vortices have been intensively studied in neutral atomic BECs where an external trapping potential such as an optical lattice and/or harmonic trap often plays an important role \cite{matthews99,madison00,aboshaeer01}. Quantized vortices can also be generated in other types of fields such as optical beams \cite{allen92,dennis09}, and these find technological applications, e.g.\ in superresolution fluorescence microscopy \cite{Klar00}.


In this paper we study vortices in a FSL by combining elements of all the above-mentioned systems. Specifically, we perform a theoretical analysis of a pair of periodically driven BECs in a double-well potential.  The periodic driving induces a synthetic gauge field in the Fock space of the system which is characterized by the boson occupation numbers of each well.  For convenience we make the assumption  that particle number is conserved, however, this is not necessary and our results are stable against some particle loss.  Due to the particle number conservation the Fock states of each BEC are uniquely identified by a single number. These numbers serve as the coordinates in a 2D FSL, akin to the site label of a 2D lattice in tight-binding models.  We show that for low energies the driven system mimics that of a non-interacting ultracold atomic gas in a rotating harmonic trap (rotating traps were an early method used to create synthetic gauge fields in BECs \cite{stringari01,sinha01}).   Within the rotating frame, the Coriolis force assumes the role of the Lorentz force experienced by a charged particle moving in a uniform magnetic field.  However, this process has inherent limitations because the rotation exerts a centrifugal force on the atoms causing the atomic cloud to expand. As the rotation speed increases there comes a point called the centrifugal limit where the cloud becomes unstable and flies apart.  In \textit{interacting} ultracold gases vortex lattices can form and the centrifugal limit marks the point where the number of vortices is comparable to the number of particles in the gas.  The gas becomes strongly correlated in this case, which is referred to as the quantum Hall regime because states analogous to fractional quantum Hall states can form.  Since our system resembles a  non-interacting rotating gas, there is no vortex lattice present. However, when the FSL version of the centrifugal limit is passed, a transition is triggered and the ground state forms a vortex.  The stability of the vortex state depends on higher-order `trapping' terms and  finite-size effects, which forces the FSL to have hard wall edges.  We quantify this transition in terms of the angular momentum in Fock space as well as the entanglement entropy between the two BECs with both showing a sudden increase in the vortex state.  

In a previous paper we predicted that vortices can occur in the Fock space of spin systems  following a sudden quench, the vortices being part of the fine structure of quantum caustics \cite{mumford19}. However, no attempt was made to specifically control the vortices. By contrast, here we suggest a scheme for generating tuneable artificial gauge potentials in Fock space that lead to controllable quantized vortices.   This adds to the arsenal of quantum simulations that can be performed in artificial dimensions.

The rest of this paper is laid out as follows: in Sec.\ \ref{Sec:Mod} we describe the basic model that is assumed throughout this paper, namely two different but mutually interacting BECs in a driven double-well potential. In Sec.\ \ref{sec:effectivehamiltonian} we start from the Floquet operator describing the periodically driven system and derive from it an effective Hamiltonian in Fock space for the total system that contains an artificial gauge potential. We compare various analytical predictions which can be made from this effective Hamiltonian against the exact Floquet theory in the main results section, Sec.\ \ref{Sec:Res}. These results include the allowed energies and angular momenta of the vortex state, density and phase plots of the wave functions in Fock space and the dependence of the vortex transition on the interactions and number of particles, and finally we show how the onset of a vortex in Fock space is associated with a jump in the entanglement entropy between the two BECs. We give our conclusions in Sec.\ \ref{sec:conclusion}. There are also two appendices which give details of two of the calculations presented in the main body of the paper.

\section{\label{Sec:Mod}Model}

The system we  investigate consists of a pair of BECs in a double-well potential undergoing time-periodic modulation.
A single BEC in a static double-well potential can be thought of as a bosonic version of the Josephson junction \cite{smerzi97}, and has been successfully realized experimentally either by trapping atoms in an actual double-well potential \cite{albiez05,levy07} or by using two different internal states of the atoms in a single-well trap \cite{zibold10}.
 In our case we assume each of the two BECs is made from a different species of atom (alternative schemes involving a single species of atom in incoherent mixtures of two different internal states in a double-well potential or even one species with four different internal states in a single-well trap are conceivable, as long as no interconversion between the two `species' takes place).    For simplicity, we will assume that there are an equal number of $N$ particles of each species present, however this is not necessary and any difference will simply cause the FSL to be rectangular in shape rather than square.  
 
 In the two-mode approximation each BEC possesses just two modes, namely the ground states associated with each well. A quantum many-particle description can then be conveniently achieved by using Schwinger's mapping onto spin-$N/2$ angular momentum operators \cite{milburn97}: $\bm{\hat{J}} = \left (\hat{J}_x,\hat{J}_y,\hat{J}_z \right )$ for one gas and $\bm{\hat{S}} = \left (\hat{S}_x,\hat{S}_y,\hat{S}_z \right )$ for the other. The labels $x,y,z$ refer to the coordinates for the abstract spaces where the spin-$N/2$ Bloch spheres are embedded rather than real spatial coordinates. Picking out the $z$ axis as the quantization axis, the state of the system is fully described in the basis of Fock states $\{\vert m,n\rangle\}$ where $\hat{J}_z \vert m,n\rangle = m\vert m,n\rangle$ and $\hat{S}_z \vert m,n\rangle = n\vert m,n\rangle$.  These states are labeled by half the particle number difference between the two wells, so they take integer values in the range $-N/2 \leq m,n\leq N/2$.  In the FSL, $m$ and $n$ label the $x$ and $y$ coordinates, respectively.  In terms of the left and right well modes the spin operators for the first BEC are $\hat{J}_x = \frac{1}{2}\left (\hat{a}_R^\dagger \hat{a}_L + \hat{a}_L^\dagger \hat{a}_R \right )$, $\hat{J}_y = -\frac{i}{2}\left (\hat{a}_R^\dagger \hat{a}_L - \hat{a}_L^\dagger \hat{a}_R \right )$ and $\hat{J}_z = \frac{1}{2}\left (\hat{a}_R^\dagger \hat{a}_R - \hat{a}_L^\dagger \hat{a}_L \right )$, where $\hat{a}_L$ ($\hat{a}^{\dag}_L$) and $\hat{a}_R$ ($\hat{a}^{\dag}_R$) annihilate (create) particles in the left and right wells, respectively, and obey the usual bosonic commutation relations. The $\bm{\hat{S}}$ operators act on the second BEC and are similarly defined, but with the operators $\hat{a}_{L}$ and $\hat{a}_{R}$  replaced by $\hat{b}_{L}$ and $\hat{b}_{R}$ etc.   In the Fock basis, the $\hat{J}_x$ and $\hat{S}_x$ operators are responsible for their respective species' transition between wells, and hence they transform a Fock state into a superposition of adjacent ones.  For example, $\hat{J}_x \vert m,n\rangle = C_+(m) \vert m+1,n\rangle + C_-(m) \vert m-1,n\rangle$ where the factors are
\begin{equation}
C_\pm(m) = \frac{1}{2}\sqrt{(N/2\mp m)(N/2\pm m+1)} \, .
\label{eq:factors}
\end{equation}

The particles from one BEC can engage in interactions with particles from either the same or the other BEC, and assuming standard short-range interactions they can only interact when they are in the same well.  This results in terms of the form $\hat{J}_z^2$ and $\hat{S}_z^2$ and $\hat{J}_z\hat{S}_z$ for intra-species and inter-species interactions, respectively \cite{mumford14}.  A positive or negative sign in front of these terms determines whether they are repulsive or attractive.

The modulation scheme we will use involves alternating between repulsive and attractive interactions between the two BECs, alongside periodic pulses controlling the tunneling of each BEC.  If the tunneling interval $\delta t$ is very short such that no other dynamics takes places over the interval, then the effect of one period of the modulation can be written in terms of the Floquet operator
\begin{equation}
\hat{U}_F = e^{-i \hat{H}_- T/4}e^{-i \hat{H}_J \delta t} e^{-i \hat{H}_+ T/2}e^{-i \hat{H}_S \delta t} e^{-i \hat{H}_- T/4} \label{eq:floq}
\end{equation}
where
\begin{eqnarray}
\hat{H}_\pm &=&- U\left ( \hat{J}_z^2 + \hat{S}_z^2 \right ) \pm W \hat{J}_z\hat{S}_z  \\
\hat{H}_J &=&- J \hat{J}_x   \\
\hat{H}_S & = & -J \hat{S}_x \ .
\end{eqnarray}
We take all of the parameters to be positive, so that $U>0$ controls the strength of \textit{attractive} intra-species interactions, whereas $W$ controls the inter-species interactions which alternate between attractive and repulsive.  The tunneling between each well is controlled by $J$ and is pulsed once for each BEC each period  for a duration $\delta t$.  We set both $U$ and $J$ to be the same for each BEC, again for convenience, but they are not required to be.  However, it will become evident that there are important values of the ratio $U/J$ that cause the ground-state wave function to change significantly, and so the ratio cannot be arbitrarily chosen.

The Floquet operator can be broken down into five steps: 1) attractive intra- and inter-species interactions for a duration of $T/4$, 2) tunneling of one BEC for a duration of $\delta t$, 3) attractive intra- and repulsive inter-species interactions for a duration $T/2$, 4) tunneling of the other BEC for a duration of $\delta t$, 5) other attractive intra- and inter-species interactions for a duration of $T/4$.  The combination of these steps results in a swirling effect in the 2D Fock space.  This effect is reminiscent of the response of electrons moving in two dimensions in response to a uniform magnetic field pointing perpendicular to the plane.  One of us (JM) has previously suggested a similar driving scheme to simulate large magnetic fields in the same 2D Fock space to investigate quantum Hall physics \cite{mumford22a}.  In this paper, our main focus will instead be on simulating weak magnetic fields involving a small fraction of a flux quantum  through each plaquette in the FSL.   

Achieving precise control over the parameters of the system is important for the physical implementation of the Floquet operator. Fortunately, the field of ultracold atoms has seen substantial progress in control techniques over the past couple of decades, and high levels of control can be accomplished.  For example, Floquet engineering of the tunneling between the wells can be achieved by modulating the lasers creating a barrier \cite{Klemmer24}, and  species-specific traps and periodic driving have been used in experiments involving ultracold atoms in double-well potentials to simulate lattice gauge theories in a setup which is quite similar to that which we are proposing here \cite{schweizer19} (see also the review \cite{aidelsburger22}). Ultracold atoms also offer the ability to control interactions between being repulsive and attractive by using a magnetic field to tune the interatomic scattering length through a Feshbach resonance. In our case we need to separately control the intra- and inter-species interactions such that the intra-species interactions are always attractive but the inter-species interactions periodically alternate in sign. This can achieved by choosing two species that each have attractive interactions within a certain range of magnetic-field magnitude,  but there is an inter-species Feshbach resonance that lies within this range. For example, certain states of $^{133}$Cs, $^{7}$Li, and $^{39}$K have considerable ranges of attractive interactions and also interspecies Feshbach resonances \cite{Pollack2009,Grobner2017,Naidon2020}. Time-dependent control of scattering lengths has also been demonstrated, e.g.\  a time-modulated magnetic field has been used to generate periodic driving of the interactions \cite{Pollack2010,Clark2017}. These types of techniques were originally used to create ultracold molecules where the interaction energy is driven at the association frequency between atoms of the same species \cite{thompson05}. More recently, and of great pertinence to the present proposal, they have also been used to control the inter-species interactions in dual species BECs of $^{87}$Rb   and  $^{41}$K    \cite{thalhammer08}, of $^{87}$Rb and $^{85}$Rb \cite{Papp2008}, of $^{168}$Yb and $^{174}$Yb  \cite{Sugawa2011}, of $^{87}$Rb and $^{133}$Cs \cite{McCarron2011}, of $^{23}$Na and $^{87}$Rb \cite{Wang2016}, of $^{87}$Rb and $^{84}$Sr and $^{87}$Rb and $^{88}$Sr \cite{Pasquiou2013}, of $^{39}$K   and  $^{87}$Rb \cite{Wacker2015}, and of $^{39}$K  and $^{41}$K \cite{tanzi18}. 

In order to observe vortices in Fock space one should be able to measure the probability distribution as a function of the coordinates $m$ and $n$ which are the number differences between the two wells for the two species. For a single species this has previously been achieved by using absorption imaging \cite{albiez05} and also resonant fluorescence detection \cite{Hume2013}. In the latter case,  resolution at the level of a single atom number was reported in a sample of 1200 atoms. To apply this technique to a dual species condensate it is sufficient that the two species have transitions which differ enough in frequency that they can be individually addressed by resonant lasers (which is often the case). It should be remembered that a single measurement of the number difference for each species will yield a single number: in order to build up a probability distribution these measurements must be repeated many times, each time on a system which has been prepared as identically as possible to all the other measurements.


\section{Effective Hamiltonian in Fock space}
\label{sec:effectivehamiltonian}

\begin{figure}[t]
\centering
\includegraphics[scale=0.6]{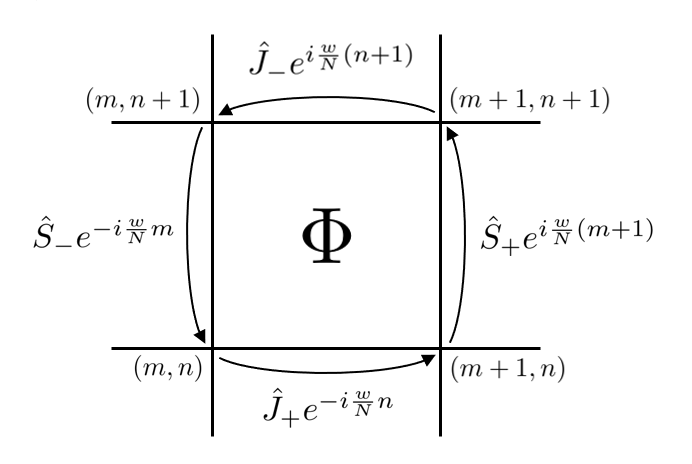}
\caption{Path taken around a plaquette in the FSL.  The magnetic flux $\Phi$  through the plaquette is defined in terms of the phase accumulated as the system completes a loop around the plaquette.  What is \textit{not} depicted in the image is the inhomogeneity in the FSL due to factors in Eq.\ \eqref{eq:factors} that come with each application of the Schwinger spin raising and lowering operators. }
\label{fig:PF}
\end{figure}

The periodic driving is capable of producing a synthetic magnetic field in the 2D Fock space formed by the eigenstates of the $\hat{J}_z$ and $\hat{S}_z$ operators.  Here, we will demonstrate the existence of the magnetic field explicitly by deriving a low-energy effective Hamiltonian with a state dependent vector potential.  To start, we note that the Floquet operator can be written in terms of an effective Hamiltonian, $\hat{U}_F = e^{-i \hat{H}_\mathrm{eff} \delta t}$.  Unfortunately, the exact form of $\hat{H}_\mathrm{eff}$ contains an infinite number of terms which can be seen by combining all of the exponentials in Eq.\ \eqref{eq:floq} using the Baker-Campbell-Hausdorff formula \cite{Sen21}.  However, for very short pulse duration, $\delta t$, the effective Hamiltonian can be well approximated by its zeroth-order term in $\delta t$, giving (a derivation is provided in Appendix \ref{app:dereff}) 
\begin{eqnarray}
\hat{H}_\mathrm{eff}/J \approx  -\frac{u}{N} \left (\hat{S}_z^2+\hat{J}_z^2 \right ) -&& \frac{1}{2} \left [ \hat{J}_+ e^{-i \frac{w}{N}\hat{S}_z}+ \hat{J}_- e^{i \frac{w}{N}\hat{S}_z} \right . \nonumber \\
&& + \left .  \hat{S}_+ e^{i \frac{w}{N}\hat{J}_z}+ \hat{S}_- e^{-i \frac{w}{N}\hat{J}_z}\right ] \nonumber \\
\label{eq:Pham}
\end{eqnarray}
where the new scaled parameters are $u=\frac{UNT}{J\delta t}$ and $w = \frac{WNT}{4}$.  We have explicitly written the Hamiltonian in terms of the Schwinger spin raising and lowering operators to highlight the phase factors which are analogous to Peierls factors arising in tight-binding models in the presence of a gauge potential. Indeed, it has been demonstrated that the terms within the square brackets of Eq.\ \eqref{eq:Pham} closely resemble the Harper-Hofstadter model and the spectrum generated from them gives rise to the celebrated Hofstadter butterfly \cite{mumford22a}.  

The synthetic magnetic flux per plaquette in the FSL can be calculated by finding the phase accumulated from one loop around a plaquette:
\begin{eqnarray}
\hat{S}_- e^{-i \frac{w}{N}\hat{J}_z} \hat{J}_- e^{i \frac{w}{N}\hat{S}_z} \hat{S}_+ e^{i \frac{w}{N}\hat{J}_z} \hat{J}_+ e^{-i \frac{w}{N}\hat{S}_z}&& \vert m,n\rangle \nonumber \\
=C_+(m)^2C_+(n)^2 e^{i \Delta \theta}&& \vert m, n\rangle 
\label{eq:plaq}
\end{eqnarray}
where $\Delta \theta = 2w/N$ is the accumulated phase.  The relation between the accumulated phase and the magnetic flux is $\Delta \theta = 2\pi \frac{\Phi}{\Phi_0}$ where $\Phi = BA$ is the magnetic flux through the plaquette and $\Phi_0 = 2 \pi$ is the magnetic flux quantum.  The FSL lattice spacing is $a = 1$, so the area of each plaquette is $A = 1$ and the magnetic field is then $B = 2w/N$.  Fig.\ \ref{fig:PF} shows each step of Eq.\ \eqref{eq:plaq} in a loop around a plaquette in the FSL.   

The synthetic magnetic-field strength can be controlled by the driving period $T$, or by the inter-species interaction strength $W$, which also plays the role of the driving amplitude.  We set $w=1$ for the remainder of the paper.  As a consequence, owing to the presence of $N$ in $w$, this implies that the driving period is short or the inter-species interactions are weak, or both.  This in turn means that the magnetic flux per plaquette becomes small, enabling us to perform a Taylor expansion of the phase factors in Eq.\ \eqref{eq:Pham}, while neglecting terms beyond quadratic order in $w/N$:
\begin{eqnarray}
\hat{H}_\mathrm{eff}/J \approx&&  -\frac{u}{N} \left (\hat{S}_z^2+\hat{J}_z^2 \right ) -  \hat{J}_x - \hat{S}_x - \frac{w}{N} \left ( \hat{J}_y \hat{S}_z -\hat{S}_y\hat{J}_z \right ) \nonumber \\
&&+ \frac{w^2}{2N^2} \left (\hat{J}_x \hat{S}_z^2+ \hat{S}_x \hat{J}_z^2 \right ) .
\label{eq:taylor}
\end{eqnarray}
The weak magnetic field allows us to further approximate Eq.\ \eqref{eq:taylor} with the continuum approximation.  This is because the magnetic length, $l_B = \sqrt{1/B} = \sqrt{N/2w}$ is much larger than the FSL lattice spacing, so eigenfunctions of the Floquet operator do not `see' the discretization of the lattice.  To make the continuum approximation, we first switch from the basis of the number difference between the two wells (i.e.\ the left and right modes) to the basis of the number difference between the even and odd modes (i.e.\  the even and odd combinations of the left and right modes) which play the distinguished role of being the single-particle ground and excited states in this two mode system.  This is accomplished with the unitary transformation $ e^{-i \frac{\pi}{2}\left (\hat{J}_y + \hat{S}_y \right ) } \hat{H}_\mathrm{eff} e^{i\frac{\pi}{2} \left (\hat{J}_y + \hat{S}_y \right )} $. For the first species this results in the spin operator transformations $\hat{J}_x \to - \hat{J}_z^\prime, \hat{J}_y \to \hat{J}_y^\prime, \hat{J}_z \to  \hat{J}_x^\prime$ where the prime indicates the operators are in the ground and excited-state basis.  Next, we perform the Holstein-Primakoff transformation \cite{holstein40}

\begin{equation}
\hat{J}_z^\prime = -N/2 + \hat{a}^\dagger \hat{a}, \hspace{20pt} \hat{J}_-^\prime = \hat{a}^\dagger \sqrt{N - \hat{a}^\dagger\hat{a}} 
\end{equation}
which maps the spin operators to a single boson mode where $\hat{a}$ annihilates a boson in that mode.  Assuming that we are dealing only with low-energy states such that $\langle \hat{a}^\dagger\hat{a}\rangle/N \ll 1$, the new spin annihilation operator can be approximated as $\hat{J}_-^\prime \approx \sqrt{N} \hat{a}^\dagger$.  The various components of the vector spin operator $\bm{\hat{J}^\prime}$ in the ground and excited state basis can likewise be approximated as $\hat{J}_x^\prime \approx \sqrt{N/2} \, \hat{x}, \hat{J}_y^\prime \approx \sqrt{N/2} \, \hat{p}_x$ and $\hat{J}_z^\prime = -N/2 + (\hat{p}_x^2+\hat{x}^2-1)/2$ where $\hat{x} = (\hat{a}^\dagger + \hat{a})/\sqrt{2}$ and $\hat{p}_x = i (\hat{a}^\dagger - \hat{a})/\sqrt{2}$ are the quadrature bosonic operators.  

The previous steps can be carried out in a similar way for the second species which is governed by the $\bm{\hat{S}^\prime}$ operators. Taking care to replace the bosonic operators by  $\hat{a} \to \hat{b}$ in intermediate steps, one obtains the results $\hat{S}_x^\prime \approx \sqrt{N/2} \, \hat{y}, \hat{S}_y^\prime \approx \sqrt{N/2} \, \hat{p}_y$ and $\hat{S}_z^\prime =- N/2 + (\hat{p}_y^2 + \hat{y}^2 - 1)/2$.  Note that we use the labels $x$ and $y$ to suggest new coordinates in Fock space. In terms of the original left and right well basis the new position operators are $\hat{x} \approx  \sqrt{2/N} \hat{J}_z$ and $\hat{y} \approx  \sqrt{2/N} \hat{S}_z$. These coordinates are not to be confused with the $x,y,z$ coordinates used to specify the original spin components $\bm{\hat{J}} = \left (\hat{J}_x,\hat{J}_y,\hat{J}_z \right )$  and $\bm{\hat{S}} = \left (\hat{S}_x,\hat{S}_y,\hat{S}_z \right )$.  Rather, we use $(x,y)$ as continuum versions of the discrete coordinates $(m,n)$ from the full quantum theory.  

In terms of these coordinates the effective Hamiltonian becomes
\begin{equation}
\hat{H}_\mathrm{eff}/J \approx \frac{1}{2} \left [  \vert \bm{\hat{p}} - \bm{\hat{A}} \vert^2 + \left (1-u\right ) \vert\bm{\hat{r}}\vert^2  \right ] 
\label{eq:hefffin}
\end{equation}
where we have omitted constant terms and terms of $\mathcal{O}\left ( 1/N\right )$.  The derivation of Eq.\ \eqref{eq:hefffin} from Eq.\ \eqref{eq:Pham} is analogous to going to the Landau-level regime in the Harper-Hofstadter model with a harmonic trap.  The position and momentum operators are $\bm{\hat{r}} = (\hat{x},\hat{y})$ and $\bm{\hat{p}} = (\hat{p}_x, \hat{p}_y)$, and $\bm{\hat{A}} = w/2 (-\hat{y}, \hat{x})$ is the vector potential operator.  For sufficiently large BECs,  $\hat{H}_\mathrm{eff}$ is the Hamiltonian of a harmonically trapped charged particle in a uniform magnetic field, $B = \bm{\nabla}\times \bm{A} = w$ pointing in the direction perpendicular to the plane of the FSL.  The missing factor of $2/N$ compared to the magnetic field calculated from Eq.\ \eqref{eq:plaq} comes from the scaling of the FSL area by $2/N$ when transforming from the Schwinger spin operators to the quadrature operators. 

 The factor of unity in the harmonic trap strength arises due to the state space being composed of many-body Fock states. Consequently, the tunneling between Fock states becomes state dependent, as indicated by the $C_\pm(m)$ factors in Eq.\ \eqref{eq:factors}.  This means that the tunneling becomes weaker as the edges are approached at $m_\mathrm{edge}, n_\mathrm{edge} = \pm N/2$ and this weakening produces the effective harmonic trap around $m = n= 0$.  The attractive intra-species interactions, characterized by strength $u$, also have harmonic dependence in Fock space but with the opposite sign and hence work against the confinement. This is because attractive intra-species interactions make it energetically favorable for the particles in each BEC to clump into one well and these states are closer to the edges of the FSL.

Equation \eqref{eq:hefffin} exhibits similarities to the single-particle Hamiltonian describing a rotating atomic gas in an harmonic trap. When expressed in a reference frame rotating with the trap this takes the form
\begin{equation}
\hat{H}_\Omega = \frac{1}{2} \left [\vert \bm{\hat{p}} - \bm{\hat{A}_\Omega} \vert^2+ (\omega_\bot^2 - \Omega^2)\vert\bm{\hat{r}}\vert^2 + \omega_\parallel^2\hat{z}^2 \right ] \ .
\end{equation}
Nonetheless, there exists one major difference that will play a crucial role in the formation of a vortex in the ground state of the system.  In rotating gas experiments it is often the goal to have the rotation frequency approach the transverse trapping frequency, $\Omega \to \omega_\bot$, in order to simulate Landau level physics.  The point where $\Omega = \omega_\bot$ is the centrifugal limit and cannot be reached because the system becomes unstable there.  In $\hat{H}_\mathrm{eff}$, the point $u = 1$ is analogous to the centrifugal limit.  In the thermodynamic limit $N \rightarrow \infty$ it marks the point where a ground-state quantum phase transition takes place from a normal phase ($u<1$) where $\langle \hat{S}_z\rangle$, $\langle \hat{J}_z \rangle = 0$ to a symmetry broken phase ($u > 1$) where $\vert \langle \hat{S}_z\rangle \vert$, $\vert \langle \hat{J}_z \rangle\vert \neq 0$.  However, for finite $N$ there are higher-order terms which we neglected in the derivation of Eq.\ \eqref{eq:hefffin} which stabilize the ground state.  Indeed, finite system sizes must stabilize the system because the FSL has hard wall edges since one cannot have more particles in a well than exist in the BEC.  Therefore, the effective full `trapping potential' in the FSL for the finite $N$ case more closely resembles the combination of a  quadratic-plus-quartic potential \cite{kavoulakis03,bretin04} and a box trap \cite{meyrath05}.

\section{\label{Sec:Res}Results}

\subsection{Eigenenergies and Eigenstates}

The spectrum and eigenfunctions of the effective Hamiltonian given in Eq.\ \eqref{eq:hefffin} can be found exactly and are expressed in terms of two quantum numbers: $n_r=0, 1, 2, \dots$ which comes from the 2D harmonic oscillator part of the Hamiltonian and is associated with the number of radial nodes in the wave function, and $m_z = 0, \pm1, \pm 2, \dots$ which is the eigenvalue of the angular momentum operator $\hat{L}_z = \hat{x}\hat{p}_y - \hat{y}\hat{p}_x$.  The spectrum is
\begin{equation}
E_{n_r,m_z} = 2\omega \left ( n_r + \vert m_z\vert/2+1/2 \right ) - w m_z/2
\label{eq:en}
\end{equation}   
where $\omega = \sqrt{1-u+w^2/4}$ is the total harmonic trap frequency which contains the additional contribution from the magnetic field, $w^2/4$.  The eigenfunction for each pair of $(n_r,m_z)$ values is
\begin{equation}
\psi_{n_r,m_z}(r,\theta) = C_{n_r,m_z} e^{im_z\theta}r^{\vert m_z\vert} e^{-\omega r^2/2} L_{n_r}^{\vert m_z\vert} (\omega r^2) 
\label{eq:wf}
\end{equation}
which is expressed in cylindrical coordinates $r = \sqrt{x^2+y^2}$ and $\theta = \arctan(y/x)$.  $C_{n_r,m_z} =\left (\frac{\omega^{\vert m_z\vert+1}n_r!}{\pi (n_r+\vert m_z\vert)!} \right )^{1/2}$ is the normalization constant and $L_a^b(x)$ is a Laguerre polynomial.  

In this paper we focus on the ground-state properties of the system, so we take $n_r = 0$ from now on.  When $u <1$, the ground-state energy is $E_{0,0} = \omega$ and the wave function is simply a Gaussian centered at $r = 0$.  In terms of the two BECs, it is centered at $m=n=0$ which is the state with an equal number of bosons of each BEC in each well.  When $u>1$, the centrifugal limit is surpassed and Eq.\ \eqref{eq:en} predicts that $E_{0,m_z} \to -\infty$ as $m_z \to \infty$. However, this is only true in the thermodynamic limit.  For finite-size BECs there can be a unique ground state with $m_z \neq 0$ which, according to Eq.\ \eqref{eq:wf}, is a vortex state with $m_z$ quanta of angular momentum.    

In order to properly compare the energy spectrum given in Eq.\ \eqref{eq:en} of the effective Hamiltonian $\hat{H}_\mathrm{eff}$  against that of the exact fully quantum system we numerically diagonalize the Floquet operator in Eq.\ \eqref{eq:floq} and find its set of eigenvalues $\{\lambda_i\}$.  From these we obtain the set $\{\epsilon_i\}$ of energies defined via the relation $ \lambda_i= e^{-i \epsilon_i \delta t }$. Because these are quasi-energies (coming from a unitary operator) they are only unique in a range of size $2\pi/\delta t$.  In our case, we only need to consider the states in this range because we chose $\delta t$ to be small enough such that $2\pi/\delta t$ is much larger than any other energy scale in the system.  

Once the energies $\{ \epsilon_i \}$ have been found for the eigenstates of the Floquet operator, the remaining challenge is to find the accompanying values of their angular momentum quantum numbers $\{ (m_z)_{i} \}$. This  can be accomplished by computing the expectation value $\langle \hat{L}_{z} \rangle$ of the angular momentum operator in each eigenstate. Reversing all of the transformations that led to the effective Hamiltonian given in Eq.\ \eqref{eq:hefffin} above, the angular momentum operator $\hat{L}_z = \hat{x}\hat{p}_y - \hat{y}\hat{p}_x$ is converted back into a many-body version in terms of the Schwinger spin operators that takes the form
\begin{equation}
\hat{L}^\mathrm{MB}_z = \frac{2}{N} \left (\hat{J}_z \hat{S}_y - \hat{S}_z\hat{J}_y \right ).
\label{eq:LZMB}
\end{equation}
$\hat{L}^\mathrm{MB}_z$ represents the angular momentum operator in the FSL and we expect its mean value $\langle \hat{L}^\mathrm{MB}_z \rangle$ to accurately give $m_z$ when computed for the low-lying energy states well away from the edge of the FSL, which are the states of primary interest in this paper. However, we note that the correspondence between $\langle \hat{L}^\mathrm{MB}_z \rangle$ and $m_{z}$ can break down for higher states due to the omission of terms that depend on the number of particles as $1/N, 1/N^{3/2},\ldots$. These terms were dropped in the derivation of $\hat{H}_\mathrm{eff}$ and lead to corrections that depend on the quadrature variables as $x^2y^2, x^2 p_y^{2}, y^{2}p_x^2,\ldots$.  As long as the states that concern us are confined to small values of the quadrature variables these corrections can be ignored, but for states that extend out to larger values the corrections become significant and mean that $\hat{L}^\mathrm{MB}_z$ is modified from the expression given above to one which no longer corresponds to the canonical angular momentum operator. 

\begin{figure}[t]
\centering
\includegraphics[scale=0.14]{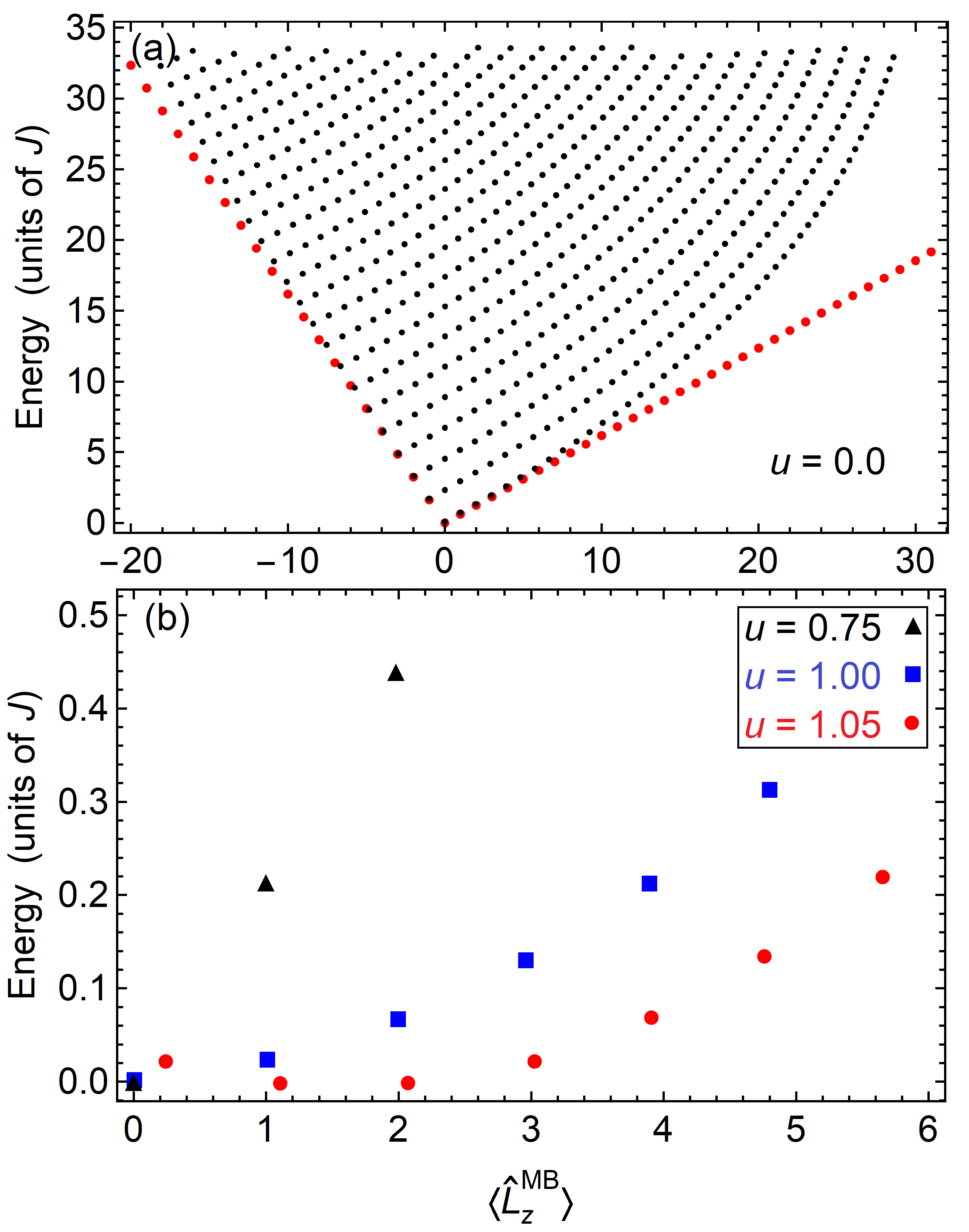}
\caption{Floquet operator energies $\epsilon_{i}$ as a function of the expectation value of the many-body angular momentum operator $\langle \hat{L}_z^\mathrm{MB} \rangle$. The black dots in panel (a) give the 500 lowest-lying energy states found numerically for $N=60$,  whereas the red dots are given by $E_{0,m_z}$ from Eq.\ \eqref{eq:en} which corresponds to the case of a harmonic trap with no higher-order trapping terms, i.e.\ assumes $N \rightarrow \infty$.  Panel (b) shows a few of the lowest states for $N=60$ for different values of $u$ where in the case of $u = 1.05$ the ground state develops nonzero angular momentum.  In both panels $w = 1$ and $\delta t = 0.01$.}
\label{fig:mz}
\end{figure}

Fig.\ \ref{fig:mz} plots the energies $\{\epsilon_{i} \}$ versus their $\langle \hat{L}_z^\mathrm{MB} \rangle$ values for a range of eigenstates of the Floquet operator  [Eq.\ \eqref{eq:floq}].  Panel (a) is generated with no intra-species interactions ($u = 0$) and shows the effect of having a finite number of particles in each BEC: the black dots are calculated numerically from the Floquet operator for $N=60$, whereas the red dots are $E_{0,m_z}$ with integer values of $m_z$ and correspond to the thermodynamic limit, $N \to \infty$, in which the effective Hamiltonian Eq.\ \eqref{eq:hefffin} becomes exact.  One can see that for the lowest-energy states $\langle \hat{L}_z^\mathrm{MB} \rangle$ gives values that agree well with integer values of $m_z$, but starting around $\langle \hat{L}_z^\mathrm{MB} \rangle = 8$ the numerical and analytic results begin to disagree. In view of the previous discussion concerning finite-size corrections these results are to be expected because states with lower angular momentum that have smaller orbits experience smaller finite-size effects, whereas those with larger angular momentum and hence larger radial extent experience larger finite size-effects. 

 The disagreement is at its extreme in the top right of panel (a),  where the energy required to increase the angular momentum by a small amount begins to diverge due to the hard wall of the 2D FSL. The approximate size of the eigenstates can be estimated from $\partial \psi_{0,m_z}/\partial r = 0$ which gives $r_\mathrm{size} \approx \sqrt{\frac{m_z}{\omega}}$.  For $u = 0$, a rough upper bound on the angular momentum which the FSL can accommodate is given by the requirement that $r_\mathrm{size} < \sqrt{N/2}$ which means $m_z^\mathrm{max} = N\omega/2$.  For the parameter values used in Fig.\ \ref{fig:mz} we obtain $\langle \hat{L}_z^\mathrm{MB} \rangle_\mathrm{max} \approx 28.6$ which is below the upper bound of $m_z^\mathrm{max} = 33$.  Panel (b) shows the lowest-energy states for different values of $u$. When $u < 1$ the ground state has zero angular momentum.  However, when $u > 1$ a dip can be seen at $\langle \hat{L}_z^\mathrm{MB} \rangle \approx 1$ indicating a ground-state transition as $u$ is changed.  Because the transition to nonzero angular momentum depends on the finite size of the system, it is \textit{not} a quantum phase transition in the strict sense that requires the thermodynamic limit be taken.

\subsection{Ground state vortex transition}

\begin{figure}[t]
\centering
\includegraphics[width=\columnwidth]{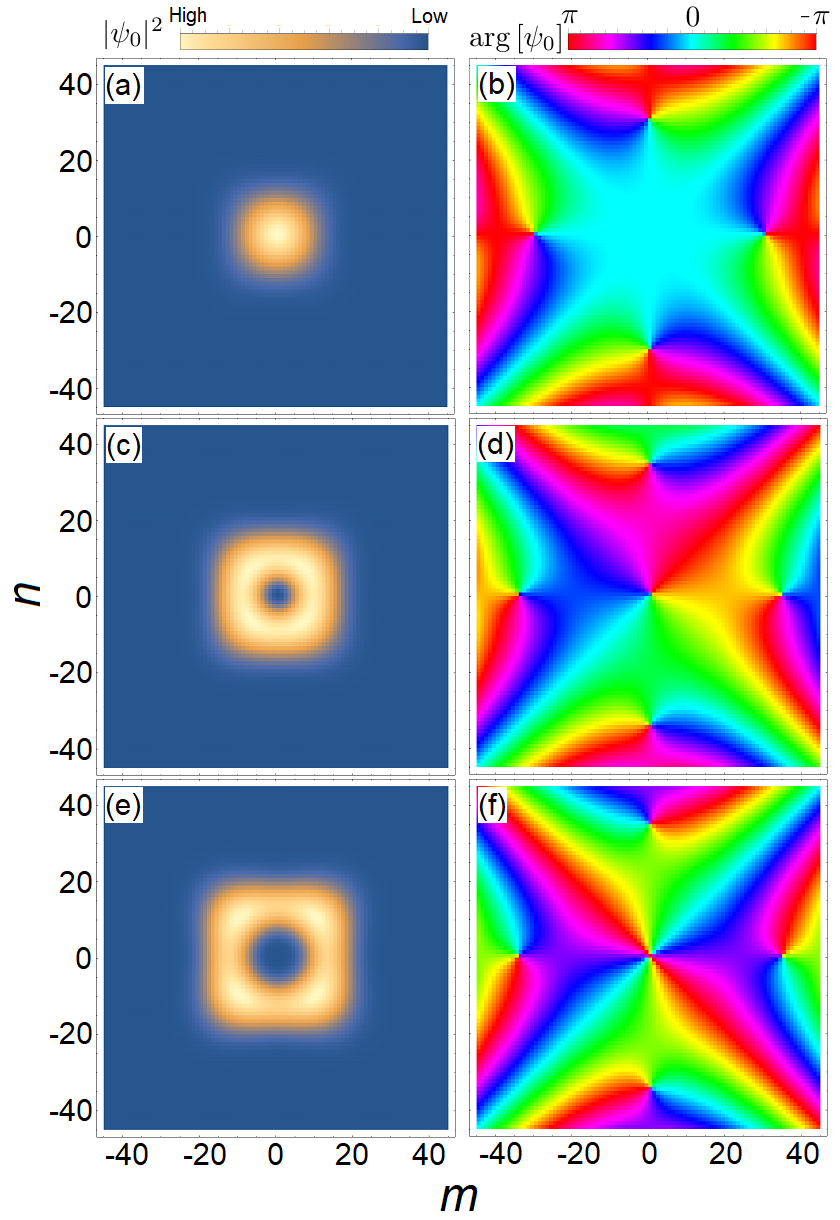}
\caption{Density and phase plots of the ground state.   Each row corresponds to a different value of $u$: $u = 1.015$ for  (a) and (b), $u = 1.020$ for (c) and (d) and $u = 1.035$ for (e) and (f).  In the phase plots (right column), as $u$ increases, the ground state transitions from no central vortex in (b) to a singly quantized one in (d), then to a doubly quantized one in (f).  Vortex cores are points of undefined phase, so the wave function must vanish at these points which results in the holes in the probability distributions in (c) and (e).  The other parameters for the plots are $w = 1$, $N = 90$, and $\delta t = 0.01$. }
\label{fig:DPG}
\end{figure}

\begin{figure}[t]
\centering
\includegraphics[scale=2.9]{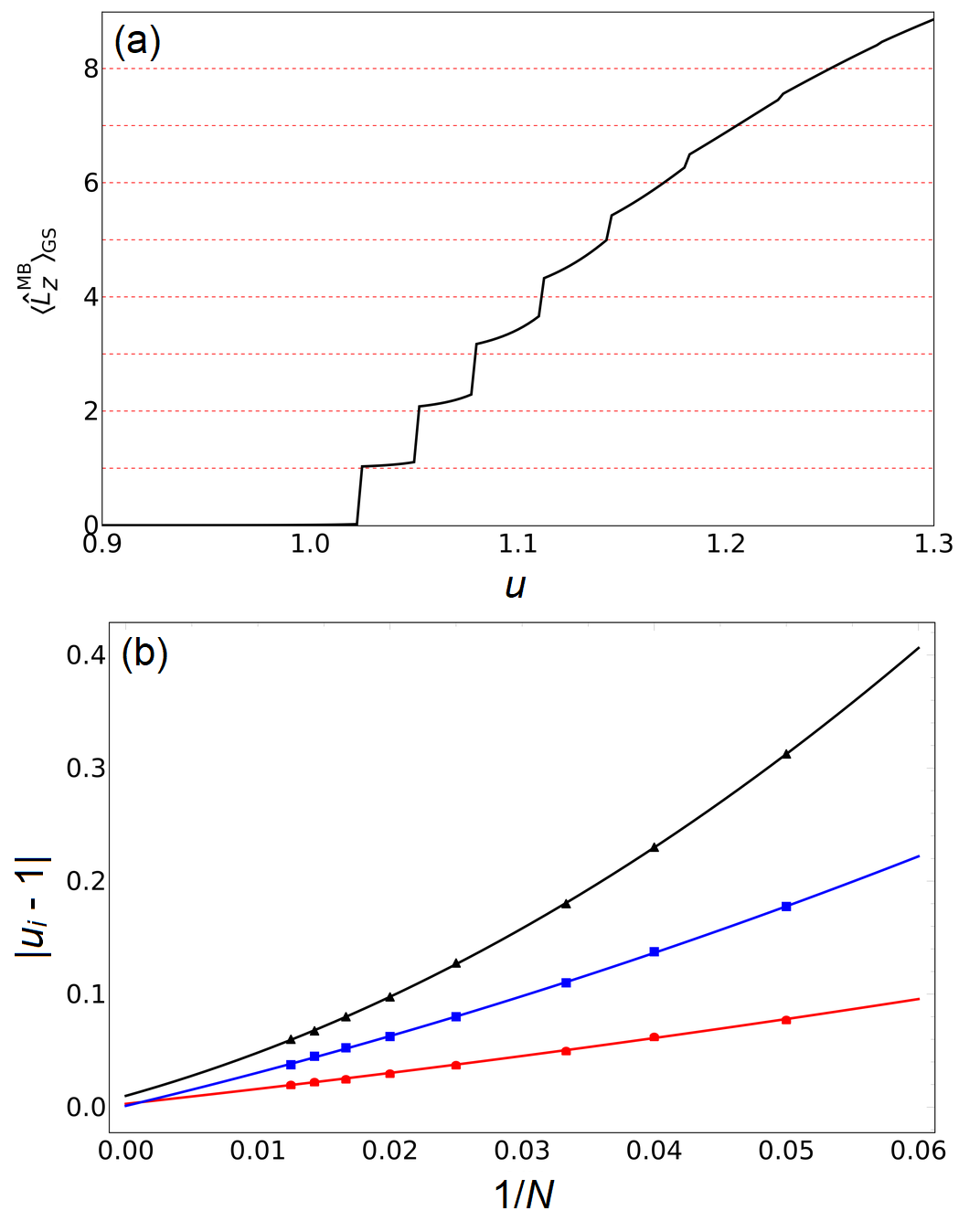}
\caption{Finite-size scaling of the vortex transitions. The black curve in panel (a) plots the ground-state expectation value of the many-body angular momentum operator $\langle \hat{L}^\mathrm{MB}_z\rangle_\mathrm{GS}$ vs $u$ which characterizes the intra-particle interaction strength, for $N = 60$. The horizontal dashed red lines show the angular momenta values $m_z$ which are the quantum numbers relevant to the solutions of $\hat{H}_{\mathrm{eff}}$.  Each step corresponds to an increase in the quantization of the central vortex.  For larger $u$, the curve becomes smoother as the quantization breaks down due to finite-size effects.  (b) $\vert u_i - 1 \vert$ as a function of $1/N$ where $u_i$, $i = 1,2,3$, are the values of $u$ where the first three steps occur and $u = 1$ is the predicted vortex transition point.  The curves are best fits of the data to polynomials of quadratic order.    The parameter values are $w = 1$ and $\delta t = 0.01$.}
\label{fig:LZG}
\end{figure}

We now turn to studying the ground state of the Floquet operator in some detail.  In Fig.\ \ref{fig:DPG} we plot both its probability distribution and its phase above and below the transition point.  First, for $u = 1.015$ in panel (a) we see that the ground state is simply a Gaussian centered at $m= n = 0$.  In panel (b), the phase at the center shows no sign of a vortex, so this state has zero angular momentum.  There are vortices near the edges of the FSL, however, the wave function is exponentially small in these regions, so they give negligible contribution to the angular momentum.  We note that for finite sizes the transition point is shifted away from $u = 1$ to a larger value which is why we get no vortex in panels (a) and (b) even though $u>1$.  When $u = 1.020$ panel (c) shows that the distribution has a hole at the center and the phase has also changed in (d), developing  a singularity at the center where the synthetic magnetic field punctures the wave function.  It is clearly shown that any circuit encircling the center once will accumulate a phase shift of $2 \pi$, so this is a singly quantized vortex.  Once more, the stability of this vortex state can be attributed to the existence of finite-size effects. These effects lead to an increase in tunneling energy towards the system's boundary resulting in confinement in Fock space that extends beyond harmonic trapping.  Finally, when $u = 1.035$ the ground state becomes more spread until its width is similar to the size of the doubly quantized vortex as seen in (e).  The phase plot in (f) confirms that a path around the central phase singularity accumulates a $4\pi$ phase shift signaling the presence of two magnetic flux quanta.  

To further quantify the transition, we plot the ground-state value of $\langle \hat{L}_z^\mathrm{MB} \rangle$ as a function of $u$ in Fig.\ \ref{fig:LZG} (a).  As $u$ is increased, the ground-state wave function undergoes a progressive widening in Fock space, such that as the width of the wave function approaches the size of the lowest angular momentum state there is a sudden quantized jump in the value of $\langle \hat{L}_z^\mathrm{MB} \rangle$. With further increments in $u$, the wave function continues to spread, approaching the size of the second lowest angular momentum state, leading to a second jump, and so on.  The red dashed horizontal lines are the integer values of the angular momentum predicted by $\hat{H}_{\mathrm{eff}}$ in Eq.\ \eqref{eq:hefffin}.  The agreement between the analytic and numerical values is best for low angular momenta, like the energies in Fig.\ \ref{fig:mz} because these states are away from the edges.  However,  as $u$ increases, eventually the width of the ground state approaches the size of the FSL and the quantization of the angular momentum breaks down as can be seen in the upper parts of the curve in Fig.\ \ref{fig:LZG} (a).   The connection between $u$ and the finite-size effects of the ground state comes from the fact that as $u$ increases the intra-species interactions become more attractive which promotes each BEC clumping into one well over the other.  Having a large number of bosons in one well over the other corresponds to large values of $\vert m\vert $ and $\vert n\vert$, thus the wave function is closer to the edge of the FSL and therefore experiences larger finite-size effects.   In the extreme example of taking the limit $u \to\infty$, we get $\hat{H}_\mathrm{eff} \to -\frac{u}{N} \left(\hat{J}_z^2 + \hat{S}_z^2 \right )$ which has degenerate ground states consisting of Fock states with $\vert m \vert = \vert n \vert = N/2$.  They will have maximum disagreement with the eigenstates of the continuum Hamiltonian in Eq.\ \eqref{eq:hefffin} which has no concept of finite size. 

In the thermodynamic limit, the ground state is predicted to possess infinite angular momentum. As a consequence, all jumps collapse into one infinitely high jump at $u = 1$. To test this prediction, in Fig.\ \ref{fig:LZG} (b) we plot the difference between the numerical value of the position of the jump and unity, $u_i -1$, as a function of $1/N$. Here, the index $i = 1, 2, 3$ corresponds to the values of $u$ where the first three jumps occur and corresponds to the bottom, middle, and top curves, respectively.  To analyze the behavior, the data of the first three jumps are fit separately  to second order polynomials which are extrapolated to $1/N = 0$.   The three curves approximately converge at the origin with values of 0.0030 (red circles, first step), 0.0012 (blue squares, second step) and 0.0100 (black triangles, third step). These results are consistent with the collapse of the first three jumps to the same transition point and we expect the same outcome for the higher-order jumps.  Thus, only in the thermodynamic does the system become unstable at $u = 1$ which is analogous to reaching the centrifugal limit of a rotating ultracold gas in an harmonic trap. 

\subsection{Entanglement entropy between the two BECs}

Quantized vortices in ordinary real space in BECs can often be accurately described within the Gross-Pitaevskii mean-field theory which does not involve entanglement. It is therefore interesting to ask whether this is also the case for Fock-space vortices or whether there is intrinsically more entanglement that is generated. On the one hand, the effective Hamiltonian in Fock space given in Eq.~(\ref{eq:hefffin}) appears to be a single particle-type Hamiltonian. On the other hand, it should be remembered that the two Fock-space coordinates $(x,y)$ each refer to different BECs and since vortices couple the two coordinates it seems that Fock-space vortices may indeed be associated with entanglement between the two underlying species of atoms. We shall therefore consider the entanglement entropy between the two BECs as another possible measure to explore in relation to the vortex transition. 

\begin{figure}[t]
\centering
\includegraphics[width=\columnwidth]{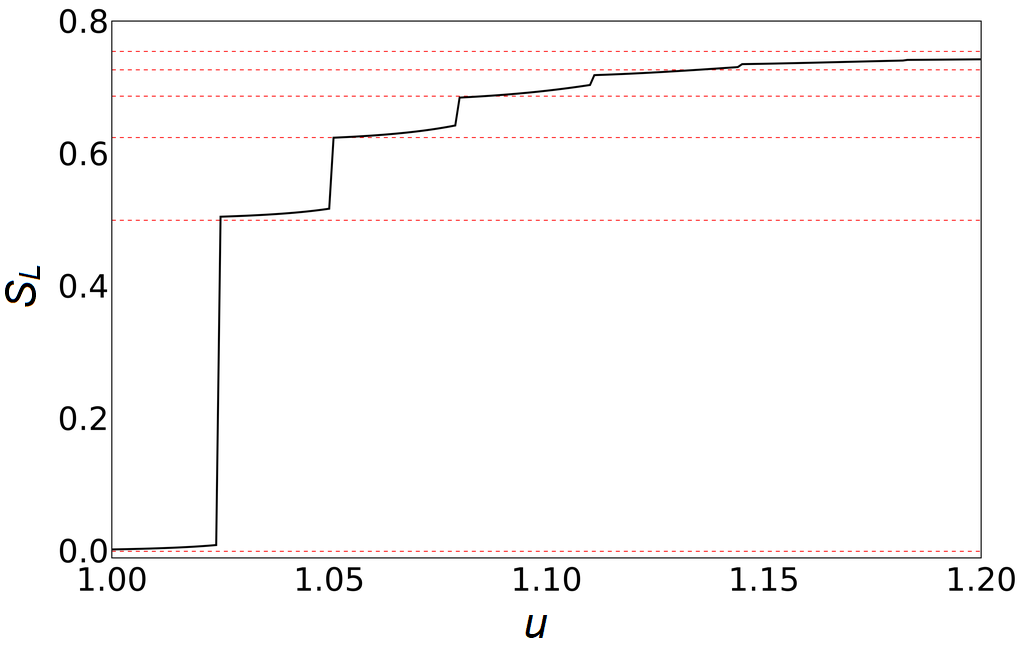}
\caption{Ground-state linear entanglement entropy between the two BECs as a function of the attractive intra-species interaction strength $u$.  The steps in the linear entropy correspond to successive transitions to higher quantized vortex states. These steps occur at the same values of $u$ as the steps seen in Fig.\ \ref{fig:LZG}. The horizontal dashed red lines are the predicted linear entanglement entropy values from Eq.\ \eqref{eq:lewf}.  The parameter values are $w = 1$, $N = 60$ and $\delta t= 0.01$.}
\label{fig:linent}
\end{figure}

For $u < 1$, when there is no central vortex present, the ground-state wave function takes the form of a 2D Gaussian located at the center of the FSL. This state is similar to that which would occur if the two BECs were not subject to driving or interacting with each other at all. Consequently, in this state, we anticipate no entanglement between the two BECs.  For $ u>1$ the two BECs must cooperate to form the vortex state and so we expect there to be some entanglement.  According to Eq.\ \eqref{eq:wf}, when $n_r = 0$, the density matrix for any eigenstate of the system is $\rho =\vert m_z \rangle \langle m_z \vert$.  In general, when one BEC is traced out, the resulting reduced density matrix, let it be $\rho_x$ corresponding only to the first BEC, will be mixed.  The mixing can be quantified in terms of the linear entropy which is related to the purity of the reduced density matrix 
\begin{equation}
S_L = 1 - \mathrm{Tr}[\rho_x^2]
\end{equation}
where $\mathrm{Tr}[\rho^2_x]$ is the purity of $\rho_x$.  The linear entanglement entropy has a minimum of $S_L = 0$ for pure states when $\mathrm{Tr}[\rho^2_x] = 1$ and a maximum of $S_L = 1-1/(N+1)$ when the reduced density matrix is completely mixed with a purity of $\mathrm{Tr}[\rho^2_x] = 1/(N+1)$. In terms of the wave functions in Eq.\ \eqref{eq:wf}, the linear entropy is (an explicit calculation can be found in Appendix \ref{app:lewf})
\begin{eqnarray}
S_L = 1 - \int dA &&\int dA^\prime\psi_{m_z}(x,y) \psi_{m_z}(x^\prime,y) ^*\nonumber \\
&&\times \psi_{m_z}(x^\prime,y^\prime) \psi_{m_z}(x,y^\prime)^*
\label{eq:lewf}
\end{eqnarray}
where the integration is taken from $-\infty$ to $\infty$ and the wave functions are in Cartesian coordinates.  We note that although we have used the reduced density matrix $\rho_x$, which signifies a trace over the $y$-coordinate BEC, we would get the same linear entropy expression if we traced out the $x$-coordinate BEC because the coupled BECs form a symmetric bipartite system.  Indeed, Eq.\ \eqref{eq:lewf} treats the $x$ and $y$ variables on equal footing.  

The first six values of the linear entropy from Eq.\ \eqref{eq:lewf} are $(m_z, S_L) = (0,0), (1,1/2), (2,5/8), (3,11/16),(4,8/11),(5,37/49)$ which are plotted as horizontal red dashed lines in Fig.\ \ref{fig:linent}.  The numerical data (black curve), calculated from the ground state of the Floquet operator, exhibits distinct jumps occurring at the same $u$ values as those depicted in Fig.\ \ref{fig:LZG}.  This confirms the idea that the entanglement entropy undergoes a discontinuity during both the first and subsequent vortex transitions.  Eventually, once again, finite system size effects become important for large $u$ and the jumps become smoother.

\section{Conclusion and Discussion}
\label{sec:conclusion}

We have shown that the periodic driving of weak interactions between two species of BECs in a double-well potential generates a weak synthetic magnetic field in the 2D FSL of the system.  Like the effect of a magnetic field on electrons in a material, the synthetic magnetic field generates vortices in the wave function in the FSL leading to states with quantized angular momentum.  With no or repulsive intra-species interactions, the ground state has zero angular momentum because the inherent confining potential in the FSL prevents the ground-state wave function from spreading to a point where it can acquire one quantum of angular momentum.   However, when the interactions are attractive and larger than $u = 1$ it becomes energetically favorable for the ground state to spread and acquire nonzero  angular momentum.  This state is only stable for a finite number of particles in each BEC.  In the thermodynamic limit, its angular momentum diverges  because higher-order terms in the confining potential, which are needed to prevent the spreading of the wave function to infinity, vanish in this limit. 

We note that the transition to a vortex state in Fock space under increasing interparticle interaction strength corresponds to a sudden jump in the occupation number difference between the original real-space wells of the double-well potential and is related to the macroscopic quantum self trapping transition that is known to occur in bosonic Josephson junctions \cite{smerzi97,albiez05}. Here it is the version that occurs for attractive interactions and hence can occur in the ground state \cite{mumford14,buonsante12,mumford2014b}.

In rotating interacting cold gases, the healing length plays a critical role in determining the size of a typical vortex, which is usually much smaller than the overall size of the gas.  This can lead to the formation of a vortex lattice \cite{aboshaeer01,schweikhard04}.  While there exist both intra- and inter-species interactions among the particles within each BEC in our system, their impact in Fock space differs. In Fock space, these interactions provide the potential energy and the synthetic magnetic field, respectively. This results in the state in Fock space exhibiting similarities to a non-interacting Bose gas, implying the absence of a healing length and therefore a vortex lattice.  What remains to shape the size of each vortex in Fock space is the `trapping potential' which has contributions from the intra-species interactions and the tunneling factors in Eq.\ \eqref{eq:factors}.  

A natural follow-up question to ask is whether interactions can occur in Fock space in order to generate an analogous quantity to a healing length and hence vortex lattices.  Although there may be many ways to simulate interactions, the most obvious approach within the context of this paper involves introducing a second pair of two-mode BECs to serve as the degrees of freedom of a second particle in the FSL. Contact interactions between each pair of BECs can simulate interactions between the two FSL particles and periodic driving within each pair ensures both particles are influenced by a synthetic magnetic field. This approach can potentially lead to strongly correlated states (in Fock space) and hence pave the way for simulating more exotic forms of matter such as fractional quantum Hall states \cite{laughlin83}.

\begin{acknowledgments}
DHJO acknowledges the support of the Natural Sciences and Engineering Research Council of Canada through Discovery Grant No. RGPIN-2017-06605.
\end{acknowledgments}

\appendix

\section{Derivation of Eq.\ \eqref{eq:Pham}\label{app:dereff}}

Starting with the Floquet operator in Eq.\ \eqref{eq:floq}, we split it into two parts

\begin{equation}
\hat{U}_F= \underbrace{e^{-i \hat{H}_- T/4}e^{-i \hat{H}_J \delta t} e^{-i \hat{H}_+ T/4}}_{\hat{U}_{F,1}}\underbrace{e^{-i \hat{H}_+ T/4}e^{-i \hat{H}_S \delta t} e^{-i \hat{H}_- T/4}}_{\hat{U}_{F,2}}.
\end{equation}
Next, we make use of the fact that $e^{i\alpha \hat{J}_z}f\left(\hat{J}_x\right)e^{-i\alpha \hat{J}_z} = f\left (\hat{J}_x \cos(\alpha) - \hat{J}_y\sin(\alpha) \right )$ (similarly for the $\hat{\bm{S}}$ operators), where $f(x)$ is a general function of $x$, to rewrite the two parts as

\begin{eqnarray}
\hat{U}_{F,1}&=& e^{-i \hat{H}_U T/4}e^{i J \left [\hat{J}_x\cos\left (W\hat{S}_zT/4\right ) -\hat{J}_y\sin\left (W\hat{S}_z T/4\right )\right ] \delta t}\nonumber \\ &&\times e^{-i \hat{H}_U T/4} \nonumber \\
\hat{U}_{F,2}&=& e^{-i \hat{H}_U T/4}e^{i J \left [\hat{S}_x\cos\left (W\hat{J}_zT/4\right ) +\hat{S}_y\sin\left (W\hat{J}_zT/4 \right )\right ] \delta t} \nonumber \\ &&\times e^{-i \hat{H}_U T/4} \nonumber \\
\end{eqnarray}
where $\hat{H}_U = - U \left ( \hat{J}_z^2+\hat{S}_z^2\right )$.  The  scaled parameters, $u= \frac{UNT}{J\delta t}$ and $w=\frac{WNT}{4}$, are substituted in and we use the Baker-Campbell-Hausdorff formula

\begin{equation}
e^{i \delta t\hat{A}} e^{i\delta t\hat{B}} = e^{i \delta t\left ( \hat{A}+\hat{B} \right ) + \frac{\left (i\delta t\right )^2}{2} \left [ \hat{A}, \hat{B} \right ]+\cdots},
\end{equation}
keeping terms up to linear order in $\delta t$.  The approximated Floquet operator is then

\begin{eqnarray}
\hat{U}_F \approx \mathrm{exp}&&\left \{ -iJ \left [-\frac{u}{N} \left (\hat{J}_z^2 + \hat{S}_z^2 \right )  - \hat{J}_x \cos\left( \frac{w\hat{S}_z}{N} \right ) \right. \right.\nonumber \\ 
&&+ \hat{J}_y \sin\left( \frac{w\hat{S}_z}{N} \right ) - \hat{S}_x \cos\left( \frac{w\hat{J}_z}{N} \right )\nonumber \\
&&\left.\left. -\hat{S}_y \sin\left( \frac{w\hat{J}_z}{N} \right )\right ]\delta t\right \}
\end{eqnarray}
and writing it as $\hat{U}_F = e^{-i \hat{H}_\mathrm{eff} \delta t}$ gives the effective Hamiltonian in Eq.\ \eqref{eq:Pham} after transforming to spin raising and lowering operators. 

\section{Derivation of Eq.\ \eqref{eq:lewf}\label{app:lewf}}

With $n_r = 0$, the density matrix for a state is $\rho = \vert m_z \rangle \langle m_z \vert$ and in the position representation is

\begin{equation}
\rho = \int dA \int dA^\prime \vert x,y \rangle \langle x^\prime, y^\prime \vert \psi_{m_z}(x,y) \psi_{m_z}(x^\prime, y^\prime)^*  
\end{equation}
where $dA = dx \, dy$ and $\psi_{m_z}(x,y) = \langle x, y \vert m_z \rangle$ is Eq.\ \eqref{eq:wf} in Cartesian coordinates.  Tracing out the $y$-coordinate gives the reduced density matrix of the $x$-coordinate BEC
\begin{equation}
\rho_x = \mathrm{Tr}_y[\rho]= \int dA\int dx^\prime dy \vert x \rangle \langle x^\prime \vert  \psi_{m_z}(x,y) \psi_{m_z}(x^\prime, y)^*
\end{equation}
and its square is 
\begin{eqnarray}
\rho_x^2 = \int dA\int dA^\prime\int dx^{\prime \prime} && \vert x \rangle \langle x^{\prime \prime}\vert \psi_{m_z}(x,y) \psi_{m_z}(x^\prime, y)^*  \nonumber \\ 
&&\times \psi_{m_z}(x^\prime,y^\prime) \psi_{m_z}(x^{\prime\prime}, y^\prime)^*. \nonumber \\
\label{eq:rho2}
\end{eqnarray}
Finally, the trace of Eq.\ \eqref{eq:rho2} will give the second term in Eq.\ \eqref{eq:lewf} in the main text.

\end{document}